\documentclass{article}
\usepackage[OT1]{fontenc} 
\usepackage{spconf,amsmath,amssymb,graphicx}
\usepackage{booktabs}
\usepackage{multirow}
\usepackage{tabularx}
\usepackage[table]{xcolor}
\usepackage{url}
\usepackage{cite}
\usepackage{float}
\usepackage{tikz}
\usetikzlibrary{arrows.meta,calc}
\usepackage{pgfplots}
\pgfplotsset{compat=1.18}

\makeatletter
\renewcommand\section{\@startsection{section}{1}{\z@}%
  {-2.2ex plus -.4ex minus -.2ex}{0.8ex plus .2ex}{\normalfont}}
\renewcommand\subsection{\@startsection{subsection}{2}{\z@}%
  {-2.0ex plus -.4ex minus -.2ex}{0.6ex plus .2ex}{\normalfont}}
\def\@tablecaptype{table}
\long\def\@makecaption#1#2{%
 \vskip 6pt
 \setbox\@tempboxa\hbox{#1. #2}%
 \ifdim \wd\@tempboxa >\hsize #1. #2\par \else \hbox
 to\hsize{\hfil\box\@tempboxa\hfil}%
 \fi
 \ifx\@captype\@tablecaptype\vskip 4pt\fi}
\g@addto@macro\small{%
  \setlength{\abovedisplayskip}{6pt plus 2pt minus 2pt}%
  \setlength{\belowdisplayskip}{6pt plus 2pt minus 2pt}%
  \setlength{\abovedisplayshortskip}{2pt plus 2pt}%
  \setlength{\belowdisplayshortskip}{4pt plus 2pt}}
\def\abstract{\par\vspace{-0.3em}\centerline{\bf ABSTRACT}\vspace{0.3em}}
\makeatother
\definecolor{accent}{HTML}{2563EB}
\definecolor{warm}{HTML}{D97706}
\definecolor{good}{HTML}{059669}
\definecolor{soft}{HTML}{6B7280}
\definecolor{ink}{HTML}{111827}

\title{SciPhi-Chat: Training a Spatial Audio Agent\\ for Question Answering, Dialogue, and Tool Use}

\name{Gijs Wijngaard\sthanks{Work done during an internship at Microsoft Research.}, Hannes Gamper, and Sebastian Braun}
\address{Microsoft Research, Redmond, WA, USA}

\begin{document}
\ninept
\maketitle

\begin{abstract}
Spatial audio question answering requires associating sound identities with locations and reasoning about their relations. This paper introduces SciPhi-Chat, a spatial audio agent that answers questions, conducts multi-turn dialogue, and calls tools on first-order ambisonic recordings. The accompanying MASU benchmark contains 12,000 samples across 42 task categories, the widest task coverage of any spatial model, with real recordings and simulated audio scenes. SciPhi-Chat achieves a score of 44.9\%, 7.4 percentage points above the previous state-of-the-art. On the external ASA-Bench, SciPhi-Chat achieves 74.5\% azimuth and 60.0\% elevation accuracy, the highest among locally evaluated models. Removing reasoning supervision reduces ASA-Bench accuracy by 17.4 percentage points. All materials will be released via \url{https://github.com/microsoft/sciphi-chat}.

\end{abstract}

\begin{keywords}
spatial audio, audio language models, tool use, dialogue, question answering
\end{keywords}

\section{Introduction}
\label{sec:intro}

Humans are capable of hearing sounds they cannot see. They can, for example, identify multiple sound sources in direction, distance, type, loudness, time, etc., and infer spatial and acoustical relations between them. Machine agents use audio-language models (ALMs) to turn audio recordings into descriptions and to answer questions. However, ALMs often use mono audio, which contains no directional information. Spatial audio requires multiple channels to encode directional cues that help ALMs associate each sound with its position. Adding the spatial dimension enables these models to describe an acoustic scene in \emph{all} human-audible acoustic dimensions.

Large language models (LLMs) can generate intermediate reasoning before answering, organizing the available information before focusing on a specific question. They can also use dialogue history to answer follow-up questions. After the conversation mentions an alarm, answering ``How far away is it?'' requires linking ``it'' to the alarm. When an answer requires a measurement or calculation, LLMs can call a tool and use its result in the response.

The initial Sci-Phi model \cite{jiangSciPhiLargeLanguage2026}, which forms the foundation of this work, describes spatial scenes and uses additional fine-tuning for targeted questions, including identifying a sound in a given direction. BAT and Spatial-Omni address spatial question answering \cite{zhengBATLearningReason2024,zhuSpatialOmniSpatialAudio2026}. Audio Dialogues and Audio Flamingo provide multi-turn conversations for training on monaural audio \cite{goelAudioDialoguesDialogues2024, kongAudioFlamingoNovel2024}. SciPhi-Chat extends these works and trains on spatial audio with text reasoning traces including follow-up references and executable tool calls.

\begin{figure*}[t]
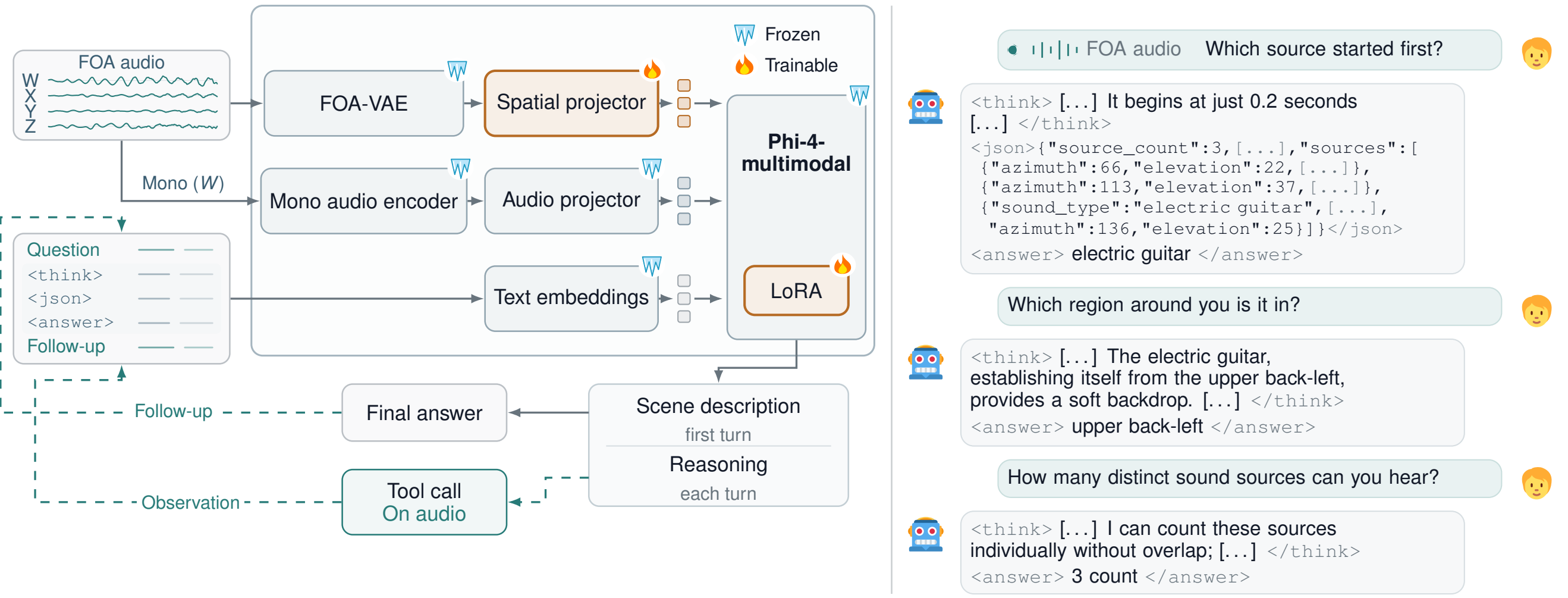

\centering
\resizebox{\textwidth}{!}{
\begingroup
\definecolor{ovblue}{HTML}{385B70}
\definecolor{ovteal}{HTML}{277B78}
\definecolor{ovamber}{HTML}{B66B24}
\definecolor{ovgray}{HTML}{68747D}
\begin{tikzpicture}[x=1mm,y=-.94mm,
  font=\sffamily\fontsize{6.8}{7.7}\selectfont,text=ink,
  box/.style={draw=ovgray!45,fill=ovgray!4,rounded corners=1.2mm,
    line width=.5pt,align=center,inner sep=1mm},
  frozen/.style={box,draw=ovblue!55,fill=ovblue!7},
  trained/.style={box,draw=ovamber,line width=.8pt,fill=ovamber!11},
  flow/.style={-{Latex[length=1.5mm,width=1.1mm]},draw=ovgray,line width=.65pt},
  loop/.style={flow,draw=ovteal,dashed},
  small/.style={font=\sffamily\fontsize{6}{6.8}\selectfont},
  token/.style={rounded corners=.25mm,line width=.4pt},
  pics/fire/.style={code={
    \fill[white] (0,0) circle (1.9);
    \fill[orange!90!red] (0,-1.65)
      .. controls (.15,-.65) and (1.25,-.55) .. (1.2,.45)
      .. controls (1.15,1.7) and (-1.25,1.7) .. (-1.2,.45)
      .. controls (-1.25,-.1) and (-.85,-.65) .. (-.65,-.9)
      .. controls (-.65,-.35) and (-.35,-.2) .. (-.25,-.05)
      .. controls (.1,-.65) and (-.15,-1.1) .. (0,-1.65);
    \fill[yellow!85!orange] (0,-.35)
      .. controls (.1,.15) and (.7,.4) .. (.6,.85)
      .. controls (.5,1.5) and (-.65,1.5) .. (-.6,.85)
      .. controls (-.55,.35) and (-.1,.15) .. (0,-.35);
  }},
  pics/icicle/.style={code={
    \fill[white] (0,0) circle (1.9);
    \path[fill=cyan!25,draw=cyan!60!ovblue,line width=.35pt,line join=round]
      (-1.35,-1.3) -- (1.35,-1.3) -- (.95,.65) -- (.55,-.5)
      -- (.05,1.65) -- (-.4,-.55) -- (-.85,1.05) -- cycle;
    \draw[white,line width=.45pt] (-.05,-.95) -- (.05,.65);
    \draw[white,line width=.4pt] (-1.05,-1) -- (-.85,.25);
  }}]
\path (0,0) -- (181,0);
\path[draw=ovblue!45,fill=ovblue!1.5,line width=.6pt,rounded corners=1.3mm]
  (30,0) rectangle (102,43);

\node[box,minimum width=25mm,minimum height=10.34mm] (recording) at (15,11) {};
\node[small,text=ovblue,inner sep=0mm] at (15,6.9) {FOA audio};
\begin{scope}[xshift=-18.15mm,yshift=-4.7mm,xscale=.65]
\input{figures/foa-waveforms}
\end{scope}
\node[box,minimum width=25mm,minimum height=15.04mm] (question) at (15,36) {};
\path[fill=ovblue!5,rounded corners=.6mm] (3.5,31.7) rectangle (26.5,40.3);
\foreach \row/\label in {30/Question,42/Follow-up} {
  \node[small,anchor=west,text=ovteal,inner sep=0mm] at (4,\row) {\label};
  \draw[ovteal!65,line width=.55pt,line cap=round] (17,\row) -- (21,\row);
  \draw[ovteal!40,line width=.55pt,line cap=round] (22.3,\row) -- (25.5,\row);
}
\foreach \row/\tag in {33/think,36/json,39/answer} {
  \node[anchor=west,text=ovblue,inner sep=0mm,
    font=\ttfamily\fontsize{6}{6.8}\selectfont] at (4,\row) {<\tag>};
  \draw[ovblue!45,line width=.55pt,line cap=round] (17,\row) -- (20.5,\row);
  \draw[ovblue!25,line width=.55pt,line cap=round] (21.8,\row) -- (25.5,\row);
}

\node[frozen,minimum width=23mm,minimum height=7.52mm] (vae) at (43,12) {FOA-VAE};
\node[frozen,minimum width=23mm,minimum height=7.52mm] (speech) at (43,24)
  {Mono audio encoder};
\node[trained,minimum width=20mm,minimum height=7.52mm] (projector) at (67,12)
  {Spatial projector};
\node[frozen,minimum width=20mm,minimum height=7.52mm] (audio) at (67,24)
  {Audio projector};
\node[frozen,minimum width=20mm,minimum height=7.52mm] (text) at (67,36)
  {Text embeddings};
\draw[flow] (recording.east |- vae.west) -- (vae.west);
\draw[flow] (recording.south) -- (15,24) -- (speech.west);
\node[small,anchor=west,text=ovblue,fill=white,inner sep=.3mm] at (17,22) {Mono (\textit{W})};
\draw[flow] (question.east) -- (text.west);
\draw[flow] (vae.east) -- (projector.west);
\draw[flow] (speech.east) -- (audio.west);
\foreach \component in {vae,speech,audio,text}
  \pic[scale=.8] at ($ (\component.north east)+(-.8,0) $) {icicle};
\pic[scale=.8] at ($ (projector.north east)+(-.8,0) $) {fire};

\foreach \y in {9.8,12,14.2}
  \draw[token,draw=ovamber,fill=ovamber!22] (79.3,\y-.7) rectangle (80.7,\y+.7);
\foreach \y in {21.8,24,26.2}
  \draw[token,draw=ovblue!65,fill=ovblue!18] (79.3,\y-.7) rectangle (80.7,\y+.7);
\foreach \y in {33.8,36,38.2}
  \draw[token,draw=ovgray!65,fill=ovgray!12] (79.3,\y-.7) rectangle (80.7,\y+.7);
\draw[flow] (projector.east) -- (78.8,12);
\draw[flow] (audio.east) -- (78.8,24);
\draw[flow] (text.east) -- (78.8,36);
\node[frozen,minimum width=16mm,minimum height=28.2mm] (decoder) at (93,26) {};
\node[align=center,font=\sffamily\bfseries\fontsize{6.8}{7.7}\selectfont] at (93,18)
  {Phi-4-\\multimodal};
\pic[scale=.8] at ($ (decoder.north east)+(-.8,0) $) {icicle};
\node[trained,minimum width=12mm,minimum height=5.64mm] (lora) at (93,35)
  {LoRA};
\pic[scale=.8] at ($ (lora.north east)+(-.8,0) $) {fire};
\foreach \y in {12,24,36}
  \draw[flow] (81.2,\y) -- (84.5,\y);

\pic[scale=.85] at (87,3.5) {icicle};
\node[small,anchor=west,inner sep=0mm] at (89.3,3.5) {Frozen};
\pic[scale=.85] at (87,7.3) {fire};
\node[small,anchor=west,inner sep=0mm] at (89.3,7.3) {Trainable};

\node[box,fill=ovblue!3,draw=ovblue!30,minimum width=30mm,minimum height=14.1mm]
  (response) at (84,54) {};
\node at (84,49.4) {Scene description};
\node[small,text=ovgray] at (84,52.7) {first turn};
\draw[ovblue!20] (71,54.1) -- (97,54.1);
\node at (84,56.6) {Reasoning};
\node[small,text=ovgray] at (84,59.9) {each turn};
\draw[flow] (decoder.south) -- (93,44.5) -- (84,44.5) -- (response.north);
\node[box,minimum width=19mm,minimum height=6.58mm] (answer) at (50,50) {Final answer};
\node[box,draw=ovteal,fill=ovteal!6,minimum width=19mm,minimum height=7.52mm]
  (tool) at (50,61) {Tool call\\\textcolor{ovteal}{On audio}};
\draw[flow] (69,50) -- (answer.east);
\draw[loop] (69,58) -- (64,58) -- (64,61) -- (tool.east);
\draw[loop] (answer.west) -- (1,50) -- (1,26) -- (15,26) -- (question.north);
\node[small,text=ovteal,fill=white,inner sep=.5mm] at (21,50) {Follow-up};
\draw[loop] (tool.west) -- (5,61) -- (5,46) -- (15,46) -- (question.south);
\node[small,text=ovteal,fill=white,inner sep=.5mm] at (23,61) {Observation};

\input{figures/conversation}
\draw[ovgray!35,line width=.5pt] (104,0) -- (104,0 |- agentthree.south);
\end{tikzpicture}
\endgroup}
\caption{SciPhi-Chat overview. Left: architecture and interaction loop with audio and conversation inputs. Right: training excerpts with shortened questions and selected reasoning/JSON fields. A follow-up locates the identified guitar behind the listener.}
\label{fig:overview}
\end{figure*}

SciPhi-Chat learns from 5.50M conversations about first-order ambisonic (FOA) recordings, which encode directional information in four audio channels. Training covers 42 task categories and includes open and multiple-choice question answering, captioning, dialogue, and tool calling. The model generates intermediate reasoning and a structured scene description before answering. The description remains in the dialogue history as context for later questions. Here, \emph{agent} refers to the model and the runtime interface: the model and the user can chat, and the model can request a tool, receive its observation, and continue responding (see Figure \ref{fig:overview}).

MASU (Massive Ambisonic Scene Understanding) evaluates spatial tasks on real recordings and synthetic data from held-out rooms. A taxonomy of spatial and acoustic tasks defines the benchmark. MASU contains 12,000 samples from real recordings and synthetic data, using datasets separate from training. Real recordings test transfer across the sim-to-real gap and simulated data test out-of-distribution scenes. SciPhi-Chat improves over the strongest current models on both real and synthetic audio. A separate evaluation tests SciPhi-Chat on the independently created ASA-Bench \cite{youWorldNotMono2026}. This work reports all four tasks in its $\mathcal{L}_1$ Spatial physics category.
This paper makes two contributions:
\begin{enumerate}\setlength{\itemsep}{0pt}\setlength{\parskip}{0pt}
\item SciPhi-Chat, a spatial audio agent trained on 42 task categories, with a shared interface for answers, follow-up dialogue, and tool calls. SciPhi-Chat describes audio segments up to 30~s containing up to 10 directional sources plus ambient background.
\item MASU, a benchmark with 12,000 samples on real and synthetic recordings, multiple answer formats, and a dialogue track. SciPhi-Chat exceeds other models by 7.4 percentage points. Further analyses evaluate the model on an external benchmark and ablate reasoning and scene-description supervision.
\end{enumerate}

\section{Related work}
\label{sec:related}
Spatial question answering uses binaural audio in BAT and OWL \cite{zhengBATLearningReason2024,biswasOWLGeometryAwareSpatial2025} and FOA in SPUR, TWNM, and Spatial-Omni \cite{sakshiSPURPlugandPlayFramework2025,youWorldNotMono2026,zhuSpatialOmniSpatialAudio2026}. FOA work also covers speech localization \cite{tangCanLargeLanguage2024} and scene description with targeted Q\&A fine-tuning in Sci-Phi \cite{jiangSciPhiLargeLanguage2026}. These spatial studies do not report training on dialogues that follow up on source references, or include executable tool calls in training. Monaural models such as Audio Flamingo train on multi-turn audio dialogue \cite{kongAudioFlamingoNovel2024}, models such as AudSemThinker and AudioThinker \cite{wijngaardAudSemThinkerEnhancingAudioLanguage2025,wuAudiothinkerGuidingLarge2026} train on intermediate reasoning, and Step-Audio 2 and Qwen3-Omni \cite{wuStepAudio2Technical2025, xuQwen3OmniTechnicalReport2025} use tool calling. While spatial models by \cite{youWorldNotMono2026} and \cite{sridharSpatialAudioQuestion2026} use reasoning, SciPhi-Chat is the first spatial audio model that combines multi-turn question answering, thinking, and tool calling.

Evaluation differs in task coverage, recording conditions, and answer formats (see also Appendix~\ref{app:benchmarks}). ASA-Bench defines 18 tasks on synthetic FOA scenes \cite{youWorldNotMono2026}. SO-Bench defines 16 subtasks and uses predominantly synthetic audio \cite{zhuSpatialOmniSpatialAudio2026}. STARSS23-QA evaluates five question types on real FOA recordings \cite{sudarsanamSpatialAudioUnderstanding2025}. MASU covers 42 categories on real recordings and simulated sound scenes, with multiple answer formats and separate single-turn and dialogue tracks. The dialogue track asks three successive questions about each recording. Figure \ref{fig:taxonomy} displays the 42 categories. A survey of 16 works found no explicit task match for nine MASU categories \cite{zhengBATLearningReason2024,biswasOWLGeometryAwareSpatial2025,tangCanLargeLanguage2024,dementyevPhaseCoderMicrophoneGeometryAgnostic2026,sakshiSPURPlugandPlayFramework2025,youWorldNotMono2026,jiangSciPhiLargeLanguage2026,zhuSpatialOmniSpatialAudio2026,liuStarBenchProbingDeep2026,sudarsanamSpatialAudioUnderstanding2025,sridharSpatialAudioMotion2025,sridharSpatialAudioQuestion2026,hyun-binSpatioTemporalAudioLanguage2026,kumarMMAUProChallengingComprehensive2026,chenSAVVYSpatialAwareness2026,devnaniLearningSpatiallyAwareLanguage2024}.
\begin{table*}[t]
\centering
\caption{MASU mean scores (\%). The task-group columns and Single use the 3,000 single-turn items, and Dialogue averages the nine groups over the 9,000 conversation turns, each answered after the model's own earlier answers. Columns run from groups whose categories require the source direction or layout (left) to groups dominated by sound identity and timing (right). Random marks the floor for uniformly random answers: one over the option count for multiple choice, one over the answer vocabulary for categorical answers, and zero for numeric answers, judged answers and captions. Bold marks the highest mean per column and underline the second highest. $^\dagger$BAT receives rendered binaural audio and only takes the first 10 seconds of audio.}
\label{tab:main}
\footnotesize
\setlength{\tabcolsep}{1pt}
\renewcommand{\arraystretch}{1.12}
\newcolumntype{Y}{>{\centering\arraybackslash}X}
\begin{tabularx}{\textwidth}{@{}c@{\hspace{6pt}}l@{\hspace{8pt}}*{9}{Y}@{\hspace{6pt}}c@{\hspace{4pt}}|c@{}}
\toprule
& & \multicolumn{9}{c}{\textbf{Task group, single-turn (\%)}} & \multicolumn{1}{c@{\hspace{4pt}}}{\multirow{2}{*}[-2pt]{\begin{tabular}{@{}c@{}}\textbf{Single}\\\textbf{Total}\end{tabular}}} & \multirow{2}{*}[-2pt]{\begin{tabular}{@{}c@{}}\textbf{Dialogue}\\\textbf{Total}\end{tabular}} \\
\cmidrule(lr){3-11}
& \textbf{Model} & \textbf{Direction} & \textbf{Ordering} & \textbf{Viewpoint} & \textbf{Relations} & \textbf{Counting} & \textbf{Reasoning} & \textbf{Room} & \textbf{Identity} & \textbf{Time} & \multicolumn{1}{c@{\hspace{4pt}}}{} & \\
\specialrule{\lightrulewidth}{\aboverulesep}{0pt}
& \textcolor{gray}{\itshape Random} & \textcolor{gray}{\itshape 19.6} & \textcolor{gray}{\itshape 25.5} & \textcolor{gray}{\itshape 29.2} & \textcolor{gray}{\itshape 21.0} & \textcolor{gray}{\itshape 17.3} & \textcolor{gray}{\itshape 5.4} & \textcolor{gray}{\itshape 14.9} & \textcolor{gray}{\itshape 13.5} & \textcolor{gray}{\itshape 26.2} & \textcolor{gray}{\itshape 19.2} & \textcolor{gray}{\itshape 18.9} \\
\specialrule{\lightrulewidth}{\aboverulesep}{0pt}
\multirow{4}{*}{\rotatebox{90}{\scriptsize\itshape Mono}} & Phi-4-mm \cite{microsoftPhi4MiniTechnicalReport2025} & 11.5 & 22.9 & 21.6 & 30.7 & 24.7 & 5.2 & 23.7 & 14.3 & 19.6 & 19.4 & 22.6 \\
& Audio Flamingo Next \cite{ghoshAudioFlamingoNext2026} & 15.1 & 25.9 & 29.4 & 44.5 & 27.4 & 25.5 & 35.9 & 38.2 & 23.3 & 29.5 & 26.4 \\
& MOSS-Audio \cite{yangMOSSAudioTechnicalReport2026} & 19.9 & 30.3 & 29.4 & 36.6 & 32.1 & 25.7 & \textbf{40.2} & \underline{41.5} & 32.4 & 32.0 & 25.1 \\
& Qwen2.5-Omni \cite{xuQwen25OmniTechnicalReport2025} & 18.6 & 32.4 & \underline{33.5} & \underline{45.5} & 34.9 & 24.3 & 20.1 & 39.7 & \underline{33.4} & 31.4 & \underline{28.2} \\
\specialrule{\lightrulewidth}{\aboverulesep}{0pt}
\multirow{4}{*}[-5pt]{\rotatebox{90}{\scriptsize\itshape Spatial}} & \textit{BAT}$^\dagger$ \cite{zhengBATLearningReason2024} & 8.5 & 13.3 & 3.1 & 32.2 & 17.1 & 6.1 & 7.0 & 11.4 & 3.6 & 11.4 & 8.6 \\
& Spatial-Omni \cite{zhuSpatialOmniSpatialAudio2026} & 20.0 & \underline{34.6} & 19.2 & 44.6 & 32.9 & 23.5 & 34.3 & 35.4 & 28.8 & 30.3 & 24.9 \\
& Sci-Phi \cite{jiangSciPhiLargeLanguage2026} + Qwen3.5-4B \cite{yangQwen3TechnicalReport2025} & \underline{32.1} & \textbf{38.0} & 31.8 & 37.4 & \textbf{48.2} & \underline{42.1} & 35.0 & 38.6 & \textbf{34.1} & \underline{37.5} & 27.9 \\
\cmidrule{2-13}
& \textbf{SciPhi-Chat (ours)} & \textbf{60.1} & 28.0 & \textbf{55.1} & \textbf{49.7} & \underline{47.9} & \textbf{45.3} & \underline{37.8} & \textbf{48.0} & 32.1 & \textbf{44.9} & \textbf{36.7} \\
\bottomrule
\end{tabularx}
\end{table*}

\section{Training the spatial audio agent}
\label{sec:method}

\subsection{Response and tool interface}
Training uses a common response format. The first assistant turn emits free-form reasoning in \texttt{<think>}, a structured scene description in \texttt{<json>} (see Appendix~\ref{app:describe}), and the final response in \texttt{<answer>}. The description lists all present sources with content description, azimuth, elevation, distance, level, and timing, plus room reverberation, volume, and background. It appears once per conversation. Later turns retain it and earlier answers in context, so training can include questions about a previously identified source.

Tools let the agent request audio measurements and use their results to answer. The model can generate a \texttt{<tool\_call>}. The runtime executes the named routine on the recording and returns its output in a message tagged \texttt{<tool\_response>}. The model then continues generation. Training computes the loss only on assistant tokens, including tool calls and subsequent answers. User prompts and tool responses are put into the context of the model, but are masked from the loss.

Four fixed signal-processing tools operate on the FOA recording. The counting and first localization tools estimate directions from $\mathbf q=\langle W[X,Y,Z]^{\mathsf T}\rangle_t\propto-\mathbf I$, where $\mathbf I$ is the active sound intensity vector \cite{merimaaSpatialImpulseResponse2004}. They select windows with low diffuseness and group their directions to estimate the source count or source directions. The second localization tool estimates a single direction from time-frequency bins that pass a direct-path dominance test \cite{nadiriLocalizationMultipleSpeakers2014}. The activity detection tool thresholds the $W$-channel level to return active time intervals. 

\subsection{Audio representation and trainable parameters}
First-order ambisonics (FOA) encodes an omnidirectional component $W$ and directional components $X$, $Y$, and $Z$ along the front-back, left-right, and up-down axes. The input uses AmbiX format, ACN channel order $[W,Y,Z,X]$ with SN3D normalization \cite{nachbarAmbixaSuggestedAmbisonics2011}. A 29.6M-parameter convolutional variational autoencoder using the backbone from \cite{braunSaladVAESemanticAudio2026} extended to 4 channels, here termed FOA-VAE, processes complex-valued four-channel spectrogram information. Pretraining optimizes FOA-VAE on the training dataset using spectral reconstruction, spatial losses and discriminators as in \cite{sudarsanamFOATokenizerLowbitrate2025}. Scene-description adaptation and agent training keep FOA-VAE frozen. FOA-VAE represents spatial audio with 512-dimensional vectors at 7.8~Hz and will be published separately in \cite{gamperSpatiallyawareVariationalAutoencoders2026}.

A three-layer projector maps the encoder's posterior statistics to spatial embeddings. In parallel, the $W$ channel passes through Phi-4-multimodal's frozen 441M-parameter Conformer speech encoder \cite{gulatiConformerConvolutionAugmentedTransformer2020}. Training zero-pads the audio to 30\,s to keep lengths the same within a batch. Spatial embeddings, audio tokens, and text condition the decoder. Agent training updates the spatial projector (12.6M parameters) and rank-16 low-rank adaptation (LoRA) modules in the decoder's attention and feed-forward projections (6.9M) \cite{huLoRALowRankAdaptation2022}. It keeps the remaining weights frozen.

\subsection{Synthetic conversations and optimization}
The audio pipeline extends the data simulation procedure from \cite{jiangSciPhiLargeLanguage2026} with additional mono source material and caption cleaning. The simulator synthesizes spatial audio scenes by convolving mono clips with FOA room impulse responses (RIRs). The RIR generator uses pyroomacoustics \cite{scheibler2018pyroomacoustics} to model shoebox rooms with sampled dimensions, surface absorption coefficients, and source and receiver positions. The scene renderer mixes the convolved clips at randomized levels and adds diffuse background audio. The recordings last from 5 to 30\,s and contain up to ten possibly temporally overlapping sources in the scene. The source clips come from Freesound, BBC Sound Effects, Sonniss, BigSoundBank, an internal Sound Effects library, Common Voice speech in 8 languages, and music from FMA and MTG-Jamendo. The pipeline cleans and post-filters sound captions using GPT-5.4 nano and TACOS \cite{primusTacosTemporallyAlignedAudio2025}.
LLMs use the resulting metadata, including source descriptions and spatial and acoustic attributes, to generate conversation traces as follows.

Figure \ref{fig:taxonomy} shows the 42-category taxonomy in 9 different groups. This taxonomy organizes the generation of questions for the generated data. Each of the 42 categories occurs in the dataset at a minimum of 0.5\% of the total data. To increase the variety of generated data, a total of 6 different LLMs, GPT-5.4-mini, GPT-4o-mini, GPT-5.1, Qwen3.5-27B, GPT-OSS-120B and Gemma-4-31B-it, each receive scene metadata, questions, and target answers and generate the \texttt{<think>} traces. The pipeline then checks each trace with regex methods for errors, including unsupported numbers, fabricated quotations, answer leakage, and invalid tool references. The LLMs regenerate rejected traces, and the pipeline checks them again. The pipeline excludes any scenes where the traces fail validation six times. The corpus contains 5.50M conversations over 1.93M scenes, with 1 to 8 conversations per scene.

A pretraining stage that predicts the structured JSON-format scene description adapts Phi-4-multimodal to the frozen FOA-VAE encoder using 10M audio-JSON training pairs, following the initial Sci-Phi recipe \cite{jiangSciPhiLargeLanguage2026}. The agentic training stage of SciPhi-Chat fine-tunes the LoRA adapters further on the conversation traces described above, using a warmup-stable-decay (WSD) learning-rate schedule \cite{huMiniCPMUnveilingPotential2024} to allow experimentation with different training durations. The schedule uses a linear warmup for 1\% of the samples. The stable phase maintains the learning rate at $10^{-5}$ and increases the global batch from 40 to 256 after 0.9M examples to lower the gradient noise without changing the learning rate \cite{smithDontDecayLearning2018}. The stable phase continues for a further 5.53M examples at global batch 256 before decay begins. SciPhi-Chat uses the checkpoint at approximately 300k examples into a linear decay that reduces the learning rate to zero over 1.5M examples at global batch 256. Fine-tuning SciPhi-Chat takes about eight B200 GPUs for approximately 4 days.

\section{MASU: tasks, recordings, and scoring}
\label{sec:benchmark}

MASU is a benchmark for spatial audio question-solving. It includes synthetic and real recordings with nine task groups and five answer formats. MASU challenges current models: the best-performing models score 70.0\% on MMAU-Pro \cite{kumarMMAUProChallengingComprehensive2026} and 70.8\% on ASA-Bench \cite{youWorldNotMono2026}, but the highest score on MASU is 44.9\%.

The benchmark contains 12,000 samples, with 163 to 462 in each of the 42 categories. The nine groups cover source identity, direction, time, room acoustics, inter-source relations, counting, ordering, higher-order reasoning, and viewpoint changes. See Figure~\ref{fig:taxonomy} for all 42 categories of questions. The same categories organize training and evaluation by task. The spatial audio taxonomy includes absent-source questions such as ``Where is the trumpet?'' when no trumpet is present. Counterfactual questions ask which sound would be loudest and where it would be if the scene omitted another source. Post-rotation layout questions ask where a source would be relative to the listener after a hypothetical turn, e.g., ``If you turned $180^\circ$, where would the water tap be?'' Onset-direction binding combines timing and localization, e.g., ``From which direction does the first sound arrive?''

The real-audio portion contains 2,005 samples (16.7\%) from 78 recording groups in STARSS23 \cite{shimadaSTARSS23AudioVisualDataset2023}, while the remaining 9,995 samples are on 2,948 scenes simulated from monaural SoundBible content and a separate set of 109 simulated rooms. Twelve categories use synthetic audio only. We rewrite or remove unclear questions, incorrect reference answers, and answer leakage. MASU has two tracks. The single-turn track has 1,171 real and 1,829 synthetic samples. The dialogue track has 9,000 samples in 3,000 three-turn conversations (278 real, 2,722 synthetic). In 4,023 of the 6,000 follow-up turns, the question refers back to exactly one earlier source without naming it, e.g., ``the sound I just asked about''.

The full benchmark contains multiple-choice QA (MCQA, 5,633 samples), categorical responses (2,349), numerical responses (728), open responses using LLM-as-a-judge (2,793), and captions (497). Multiple-choice and categorical answers use option or vocabulary matching. Questions that include numeric answers use appropriately rounded values and error tolerances. GPT-5.4-mini assigns rubric-based scores to open answers and extracts assertions from captions for source-direction pair F1 scoring. Closed-form answers score one or zero, open answers receive the judge score, and captions the pair F1. Each task group score averages its item scores, and each track's score averages the nine groups.

\begin{figure}[t]
\centering
\definecolor{taxA}{HTML}{2A78D6}
\definecolor{taxB}{HTML}{EB6834}
\definecolor{taxC}{HTML}{1BAF7A}
\definecolor{taxD}{HTML}{EDA100}
\definecolor{taxE}{HTML}{E87BA4}
\definecolor{taxF}{HTML}{008300}
\definecolor{taxG}{HTML}{4A3AA7}
\definecolor{taxH}{HTML}{E34948}
\definecolor{taxI}{HTML}{7C3A8F}
\begin{tikzpicture}[x=1mm,y=-1mm,
  tile/.style={rounded corners=0.35mm,inner sep=0.25mm,align=center,font=\fontsize{5.4}{5.9}\selectfont,anchor=center},
  hdr/.style={anchor=base west,inner sep=0,font=\fontsize{6.4}{7}\selectfont\bfseries},
  badge/.style={anchor=north east,inner sep=0,font=\fontsize{4.6}{4.6}\selectfont,text=black!70}]
\node[hdr] at (0.10,2.20) {Source identity};
\fill[taxA] (0.00,2.65) rectangle (21.09,2.95);
\path[rounded corners=0.35mm,fill=taxA!40] (0.00,3.10) rectangle (10.19,9.90);
\node[tile,text width=9.59mm] at (5.10,6.80) {\hspace{0pt}sound class};
\node[badge] at (9.89,3.35) {7};
\path[rounded corners=0.35mm,fill=taxA!40] (10.54,3.10) rectangle (21.09,9.90);
\node[tile,text width=9.94mm] at (15.81,6.80) {\hspace{0pt}identity by attribute};
\node[badge] at (20.79,3.35) {7};
\path[rounded corners=0.35mm,fill=taxA!40] (0.00,10.25) rectangle (10.19,17.05);
\node[tile,text width=9.59mm] at (5.10,13.95) {\hspace{0pt}speech content};
\node[badge] at (9.89,10.50) {5};
\path[rounded corners=0.35mm,fill=taxA!15] (10.54,10.25) rectangle (21.09,17.05);
\node[tile,text width=9.94mm] at (15.81,13.95) {\hspace{0pt}scene caption};
\node[badge] at (20.79,10.50) {1};
\node[hdr] at (21.99,2.20) {Direction};
\fill[taxB] (21.89,2.65) rectangle (63.29,2.95);
\path[rounded corners=0.35mm,fill=taxB!40] (21.89,3.10) rectangle (35.34,9.90);
\node[tile,text width=12.85mm] at (28.61,6.80) {\hspace{0pt}azimuth};
\node[badge] at (35.04,3.35) {12};
\path[rounded corners=0.35mm,fill=taxB!40] (35.69,3.10) rectangle (49.14,9.90);
\node[tile,text width=12.85mm] at (42.41,6.80) {\hspace{0pt}elevation};
\node[badge] at (48.84,3.35) {11};
\path[rounded corners=0.35mm,fill=taxB!40] (49.49,3.10) rectangle (63.29,9.90);
\node[tile,text width=13.20mm] at (56.39,6.80) {\hspace{0pt}distance};
\node[badge] at (62.99,3.35) {9};
\path[rounded corners=0.35mm,fill=taxB!40] (21.89,10.25) rectangle (42.24,17.05);
\node[tile,text width=19.75mm] at (32.06,13.95) {\hspace{0pt}full region};
\node[badge] at (41.94,10.50) {6};
\path[rounded corners=0.35mm,fill=taxB!40] (42.59,10.25) rectangle (63.29,17.05);
\node[tile,text width=20.10mm] at (52.94,13.95) {\hspace{0pt}hemisphere};
\node[badge] at (62.99,10.50) {6};
\node[hdr] at (64.19,2.20) {Time};
\fill[taxC] (64.09,2.65) rectangle (86.00,2.95);
\path[rounded corners=0.35mm,fill=taxC!22] (64.09,3.10) rectangle (74.70,9.90);
\node[tile,text width=10.00mm] at (69.40,6.80) {\hspace{0pt}onset/\hspace{0pt}offset};
\node[badge] at (74.40,3.35) {2};
\path[rounded corners=0.35mm,fill=taxC!22] (75.05,3.10) rectangle (86.00,9.90);
\node[tile,text width=10.35mm] at (80.52,6.80) {\hspace{0pt}temporal order};
\node[badge] at (85.70,3.35) {2};
\path[rounded corners=0.35mm,fill=taxC!22] (64.09,10.25) rectangle (86.00,17.05);
\node[tile,text width=21.31mm] at (75.05,13.95) {\hspace{0pt}onset--\hspace{0pt}direction bind};
\node[badge] at (85.70,10.50) {2};
\node[hdr] at (0.10,20.35) {Room acoustics};
\fill[taxD] (0.00,20.80) rectangle (38.88,21.10);
\path[rounded corners=0.35mm,fill=taxD!15] (0.00,21.25) rectangle (12.61,28.05);
\node[tile,text width=12.01mm] at (6.30,24.95) {\hspace{0pt}reverb amount};
\node[badge] at (12.31,21.50) {1};
\path[rounded corners=0.35mm,fill=white,draw=taxD,line width=0.5pt] (12.96,21.25) rectangle (25.57,28.05);
\node[tile,text width=12.01mm] at (19.26,24.95) {\hspace{0pt}reverb time};
\node[badge] at (25.27,21.50) {\textbf{0}};
\path[rounded corners=0.35mm,fill=taxD!15] (25.92,21.25) rectangle (38.88,28.05);
\node[tile,text width=12.36mm] at (32.40,24.95) {\hspace{0pt}room size};
\node[badge] at (38.58,21.50) {1};
\path[rounded corners=0.35mm,fill=white,draw=taxD,line width=0.5pt] (0.00,28.40) rectangle (12.61,35.20);
\node[tile,text width=12.01mm] at (6.30,32.10) {\hspace{0pt}loudness};
\node[badge] at (12.31,28.65) {\textbf{0}};
\path[rounded corners=0.35mm,fill=white,draw=taxD,line width=0.5pt] (12.96,28.40) rectangle (25.57,35.20);
\node[tile,text width=12.01mm] at (19.26,32.10) {\hspace{0pt}background};
\node[badge] at (25.27,28.65) {\textbf{0}};
\path[rounded corners=0.35mm,fill=white,draw=taxD,line width=0.5pt] (25.92,28.40) rectangle (38.88,35.20);
\node[tile,text width=12.36mm] at (32.40,32.10) {\hspace{0pt}indoor/\hspace{0pt}outdoor};
\node[badge] at (38.58,28.65) {\textbf{0}};
\node[hdr] at (39.78,20.35) {Inter-source relations};
\fill[taxE] (39.68,20.80) rectangle (86.00,21.10);
\path[rounded corners=0.35mm,fill=taxE!40] (39.68,21.25) rectangle (50.91,28.05);
\node[tile,text width=10.63mm] at (45.29,24.95) {\hspace{0pt}relative direction};
\node[badge] at (50.61,21.50) {7};
\path[rounded corners=0.35mm,fill=taxE!40] (51.26,21.25) rectangle (62.49,28.05);
\node[tile,text width=10.63mm] at (56.87,24.95) {\hspace{0pt}relative distance};
\node[badge] at (62.19,21.50) {8};
\path[rounded corners=0.35mm,fill=white,draw=taxE,line width=0.5pt] (62.84,21.25) rectangle (74.07,28.05);
\node[tile,text width=10.63mm] at (68.45,24.95) {\hspace{0pt}relative loudness};
\node[badge] at (73.77,21.50) {\textbf{0}};
\path[rounded corners=0.35mm,fill=taxE!15] (74.42,21.25) rectangle (86.00,28.05);
\node[tile,text width=10.98mm] at (80.21,24.95) {\hspace{0pt}angular separation};
\node[badge] at (85.70,21.50) {1};
\path[rounded corners=0.35mm,fill=white,draw=taxE,line width=0.5pt] (39.68,28.40) rectangle (50.91,35.20);
\node[tile,text width=10.63mm] at (45.29,32.10) {\hspace{0pt}colocation};
\node[badge] at (50.61,28.65) {\textbf{0}};
\path[rounded corners=0.35mm,fill=white,draw=taxE,line width=0.5pt] (51.26,28.40) rectangle (62.49,35.20);
\node[tile,text width=10.63mm] at (56.87,32.10) {\hspace{0pt}in between};
\node[badge] at (62.19,28.65) {\textbf{0}};
\path[rounded corners=0.35mm,fill=taxE!29] (62.84,28.40) rectangle (74.07,35.20);
\node[tile,text width=10.63mm] at (68.45,32.10) {\hspace{0pt}same attribute};
\node[badge] at (73.77,28.65) {3};
\path[rounded corners=0.35mm,fill=taxE!40] (74.42,28.40) rectangle (86.00,35.20);
\node[tile,text width=10.98mm] at (80.21,32.10) {\hspace{0pt}identity--\hspace{0pt}location bind};
\node[badge] at (85.70,28.65) {5};
\node[hdr] at (0.10,38.50) {Counting};
\fill[taxF] (0.00,38.95) rectangle (18.22,39.25);
\path[rounded corners=0.35mm,fill=taxF!22] (0.00,39.40) rectangle (8.76,46.20);
\node[tile,text width=8.16mm] at (4.38,43.10) {\hspace{0pt}source count};
\node[badge] at (8.46,39.65) {2};
\path[rounded corners=0.35mm,fill=white,draw=taxF,line width=0.5pt] (9.11,39.40) rectangle (18.22,46.20);
\node[tile,text width=8.51mm] at (13.67,43.10) {\hspace{0pt}count in region};
\node[badge] at (17.92,39.65) {\textbf{0}};
\path[rounded corners=0.35mm,fill=taxF!29] (0.00,46.55) rectangle (8.76,53.35);
\node[tile,text width=8.16mm] at (4.38,50.25) {\hspace{0pt}region occupancy};
\node[badge] at (8.46,46.80) {3};
\path[rounded corners=0.35mm,fill=taxF!29] (9.11,46.55) rectangle (18.22,53.35);
\node[tile,text width=8.51mm] at (13.67,50.25) {\hspace{0pt}enumerate + location};
\node[badge] at (17.92,46.80) {3};
\node[hdr] at (19.12,38.50) {Ordering};
\fill[taxG] (19.02,38.95) rectangle (36.40,39.25);
\path[rounded corners=0.35mm,fill=taxG!15] (19.02,39.40) rectangle (27.36,46.20);
\node[tile,text width=7.74mm] at (23.19,43.10) {\hspace{0pt}spatial order};
\node[badge] at (27.06,39.65) {1};
\path[rounded corners=0.35mm,fill=taxG!36] (27.71,39.40) rectangle (36.40,46.20);
\node[tile,text width=8.09mm] at (32.06,43.10) {\hspace{0pt}extremum source};
\node[badge] at (36.10,39.65) {4};
\path[rounded corners=0.35mm,fill=taxG!15] (19.02,46.55) rectangle (36.40,53.35);
\node[tile,text width=16.78mm] at (27.71,50.25) {\hspace{0pt}saliency direction};
\node[badge] at (36.10,46.80) {1};
\node[hdr] at (37.30,38.50) {Higher-order reasoning};
\fill[taxH] (37.20,38.95) rectangle (66.77,39.25);
\path[rounded corners=0.35mm,fill=taxH!15] (37.20,39.40) rectangle (46.71,46.20);
\node[tile,text width=8.91mm] at (41.95,43.10) {\hspace{0pt}counterfactual};
\node[badge] at (46.41,39.65) {1};
\path[rounded corners=0.35mm,fill=taxH!29] (47.06,39.40) rectangle (56.57,46.20);
\node[tile,text width=8.91mm] at (51.81,43.10) {\hspace{0pt}multi-hop};
\node[badge] at (56.27,39.65) {3};
\path[rounded corners=0.35mm,fill=taxH!15] (56.92,39.40) rectangle (66.77,46.20);
\node[tile,text width=9.26mm] at (61.85,43.10) {\hspace{0pt}physical consistency};
\node[badge] at (66.47,39.65) {1};
\path[rounded corners=0.35mm,fill=taxH!22] (37.20,46.55) rectangle (46.71,53.35);
\node[tile,text width=8.91mm] at (41.95,50.25) {\hspace{0pt}scene abduction};
\node[badge] at (46.41,46.80) {2};
\path[rounded corners=0.35mm,fill=taxH!29] (47.06,46.55) rectangle (56.57,53.35);
\node[tile,text width=8.91mm] at (51.81,50.25) {\hspace{0pt}spatial caption};
\node[badge] at (56.27,46.80) {3};
\path[rounded corners=0.35mm,fill=taxH!15] (56.92,46.55) rectangle (66.77,53.35);
\node[tile,text width=9.26mm] at (61.85,50.25) {\hspace{0pt}absent source};
\node[badge] at (66.47,46.80) {1};
\node[hdr] at (67.67,38.50) {Viewpoint changes};
\fill[taxI] (67.57,38.95) rectangle (86.00,39.25);
\path[rounded corners=0.35mm,fill=taxI!22] (67.57,39.40) rectangle (76.44,46.20);
\node[tile,text width=8.26mm] at (72.01,43.10) {\hspace{0pt}allocentric reframe};
\node[badge] at (76.14,39.65) {2};
\path[rounded corners=0.35mm,fill=taxI!22] (76.79,39.40) rectangle (86.00,46.20);
\node[tile,text width=8.61mm] at (81.39,43.10) {\hspace{0pt}post-rotation layout};
\node[badge] at (85.70,39.65) {2};
\path[rounded corners=0.35mm,fill=white,draw=taxI,line width=0.5pt] (67.57,46.55) rectangle (86.00,53.35);
\node[tile,text width=17.83mm] at (76.79,50.25) {\hspace{0pt}steering to source};
\node[badge] at (85.70,46.80) {\textbf{0}};
\end{tikzpicture}
\caption{Spatial audio taxonomy for the MASU benchmark. Each category has 163 to 462 samples: 43 to 84 in the single-turn track and 119 to 379 in the dialogue track. Numbers count explicit question or caption-task coverage among 16 audited spatial-audio works. Outlines mark the nine categories with no explicit task match in that set.}
\label{fig:taxonomy}
\end{figure}
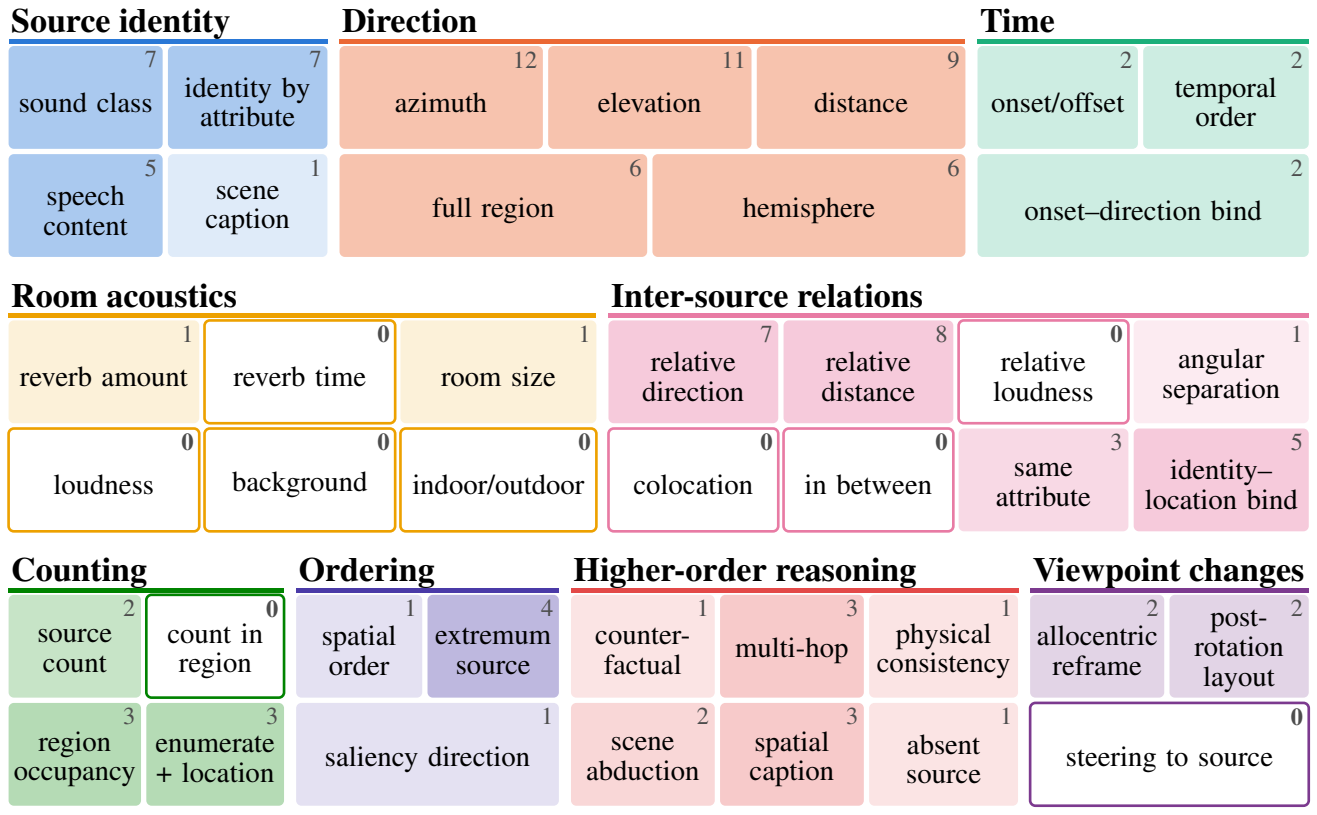
\begin{table}[t]
\centering
\caption{ASA-Bench accuracy (\%) on all four tasks in the \textit{$\mathcal{L}_1$ Spatial physics} category. Acoustics denotes environment acoustics (55 questions per task, 220 total). All four tasks use four-option multiple choice. Local scores use exact parsing with an embedding fallback. Bold marks the highest score shown per column and underline the second highest. $^\dagger$ASA-Bench reports these results: BAT-SFT uses a semantic-answer audit and TWNM-SAPO uses exact MCQA matching \cite{youWorldNotMono2026}.}
\label{tab:external}
\footnotesize
\setlength{\tabcolsep}{1pt}
\begin{tabular*}{\columnwidth}{@{}c@{\hspace{4pt}\extracolsep{\fill}}lcccc@{}}
\toprule
& \textbf{Model} & \textbf{Azimuth} & \textbf{Elevation} & \textbf{Distance} & \textbf{Acoustics} \\
\midrule
& \textcolor{gray}{\itshape Random} & \textcolor{gray}{\itshape 25.0} & \textcolor{gray}{\itshape 25.0} & \textcolor{gray}{\itshape 25.0} & \textcolor{gray}{\itshape 25.0} \\
\midrule
\multirow{4}{*}{\rotatebox{90}{\scriptsize\itshape Mono}} & Phi-4-mm \cite{microsoftPhi4MiniTechnicalReport2025} & 47.3 & \underline{58.2} & \underline{61.8} & 67.3 \\
& Audio Flamingo Next \cite{ghoshAudioFlamingoNext2026} & 40.0 & 52.7 & \underline{61.8} & 74.5 \\
& MOSS-Audio \cite{yangMOSSAudioTechnicalReport2026} & 43.6 & 41.8 & 40.0 & 58.2 \\
& Qwen2.5-Omni \cite{xuQwen25OmniTechnicalReport2025} & 25.5 & 43.6 & 45.5 & 54.5 \\
\midrule
\multirow{5}{*}{\rotatebox{90}{\scriptsize\itshape Spatial}} & BAT-SFT$^\dagger$ \cite{zhengBATLearningReason2024} & 27.3 & 47.3 & 25.5 & 7.3 \\
& Sci-Phi \cite{jiangSciPhiLargeLanguage2026} + Qwen3.5-4B \cite{yangQwen3TechnicalReport2025} & 23.6 & 43.6 & 14.5 & \underline{80.0} \\
& Spatial-Omni \cite{zhuSpatialOmniSpatialAudio2026} & 27.3 & 40.0 & 52.7 & 41.8 \\
& TWNM-SAPO$^\dagger$ \cite{youWorldNotMono2026} & \underline{54.5} & 47.3 & \textbf{70.9} & 69.1 \\
& \textbf{SciPhi-Chat (ours)} & \textbf{74.5} & \textbf{60.0} & 47.3 & \textbf{85.5} \\
\bottomrule
\end{tabular*}
\end{table}

\section{Experiments}
\label{sec:exp}

Table~\ref{tab:main} compares SciPhi-Chat with seven other models on MASU. In the dialogue track, each model sees its own earlier answers as chat turns (as a text transcript for Spatial-Omni and BAT). BAT receives HRTF-rendered binaural audio truncated or padded to 10\,s. Spatial-Omni and the Sci-Phi combination receive FOA. The four monaural models receive the omnidirectional ($W$) channel. Table~\ref{tab:external} reports ASA-Bench's $\mathcal{L}_1$ Spatial physics category and uses the benchmark authors' BAT-SFT results. The Sci-Phi + Qwen3.5-4B combination supplies the JSON-format full scene description from Sci-Phi to Qwen3.5-4B, both frozen. The Qwen3.5-4B model without audio input answers the benchmark question from the description of Sci-Phi. Qwen3.5-4B's size approximates SciPhi-Chat's language backbone \cite{jiangSciPhiLargeLanguage2026,qwenteamQwen35OmniTechnicalReport2026}, while being a more modern model with improved reasoning capability.

SciPhi-Chat obtains the highest single-turn score in Table \ref{tab:main}: 44.9\%, compared with 37.5\% for Sci-Phi + Qwen3.5-4B and 30.3\% for Spatial-Omni, a gain of 7.4 percentage points over the previous best.
SciPhi-Chat leads Spatial-Omni in eight of the nine task groups, all except ordering, and the combination in six. SciPhi-Chat scores 60.1\% in direction, compared to 32.1\% for the combination and 20.0\% for Spatial-Omni. Within this group, SciPhi-Chat scores 67.1\% on elevation, versus 31.6\% for the combination and 25.3\% for Spatial-Omni. In the dialogue track, SciPhi-Chat scores 36.7\%, compared with 28.2\% for the strongest baseline, Qwen2.5-Omni. It leads six of the nine task groups and follows closely in ordering, identity, and time.

In ASA-Bench, Table \ref{tab:external} reports all four tasks in the $\mathcal{L}_1$ Spatial physics category. SciPhi-Chat exceeds BAT and Sci-Phi + Qwen3.5-4B on all four tasks. It exceeds Spatial-Omni in azimuth, elevation, and environment acoustics, but scores lower in distance. Its environment acoustics accuracy is 85.5\%, compared to 69.1\% reported by the authors for TWNM-SAPO \cite{youWorldNotMono2026}.

The monaural models receive the omnidirectional $W$ channel, which discards directional cues (see Appendix~\ref{app:channels}). On MASU, they score near or below the random floor in the direction group, between 11.5\% and 19.9\% against 19.6\%, and Phi-4-mm scores 19.4\% on the single-turn track against a floor of 19.2\%. On ASA-Bench, always choosing answer option B on the four-option multiple-choice questions scores 49.2\%, although the evaluation found no models using this shortcut. Also, the performance of mono models being above-chance on spatial attributes indicates possible answer leakage in questions.
BAT's training uses 10-second binaural scenes with one or two sources and fixed answer templates \cite{zhengBATLearningReason2024}. MASU has a larger discrepancy between the strongest spatial systems and the monaural models than ASA-Bench's spatial-physics subset. 

Other experiments include training separate models without reasoning or without both reasoning and scene descriptions, using the same training procedure. Each model answers in its training response format during the evaluation.
The evaluated checkpoints differ more in performance on ASA-Bench than on an earlier version of MASU.
Training without \texttt{<think>} blocks retains scene-description and answer supervision. For ASA-Bench, the non-think model scores 39.9\%, versus 57.3\% for an earlier full-format checkpoint, a difference of 17.4 percentage points. The corresponding MASU scores for those without \texttt{<think>} are 43.07\%, against 44.95\% for the full model, a change of 1.88 percentage points. Answer-only training removes both \texttt{<think>} and \texttt{<json>} blocks, and scores 35.9\% in ASA-Bench versus 56.0\% for the matching full-format model, a difference of 20.1 percentage points, while MASU scores are 46.69\% and 47.07\%, a change of 0.38 percentage points. MASU shares the taxonomy of the training tasks, whereas ASA-Bench uses independently defined questions \cite{youWorldNotMono2026}. One hypothesis for this discrepancy between benchmarks is that reasoning supervision supports transfer across question formulations to allow the model to generalize better to a different benchmark.

In preliminary experiments on an earlier version of MASU, forcing the model to always call tools reduces accuracy by 5.1 percentage points on average. Constraining tool answers only on specific question types raises performance (see Appendix~\ref{app:tools}). Forcing tools also helps in specific MASU categories: for example, the time-frequency-based localization tool improves source counting from 46.0\% to 65.1\% and the activity detection tool improves the accuracy in direction of first sound from 56.7\% to 73.2\%. When the model calls the first localization tool itself, accuracy rises from 56.7\% to 67.0\%. Overall, the combination of signal processing analysis with the model's agentic reasoning has potential but requires further research to improve tool selection and integration.

\section{Conclusions}
\label{sec:discussion}
SciPhi-Chat combines spatial question answering, follow-up dialogue, and tool calling in a shared training format. On MASU, covering 42 task categories in real recordings and synthetic scenes, it scores 44.9\%, 7.4 percentage points above the strongest baseline evaluated. MASU's real recordings test whether training on simulated scenes transfers to recorded acoustic environments. SciPhi-Chat also leads over state-of-the-art systems on ASA-Bench's azimuth and elevation tasks.

Removing reasoning supervision reduces accuracy more on ASA-Bench than on MASU. This possibly indicates that reasoning improves performance on questions from other taxonomies, out-of-distribution from the domains our work focuses on. Separately, forced tool observations help selected tasks but reduce overall accuracy, which leaves tool selection unresolved. Future work will explore tool selection and put more emphasis on longer context with longer multi-turn conversations including multi-audio.

\par\pagebreak

\bibliographystyle{IEEEbib-abbrev}
\bibliography{bibliography}

\appendix
\raggedbottom
\section*{\centering\bfseries APPENDIX}
\setcounter{subsection}{0}
\renewcommand{\thesubsection}{\Alph{subsection}}

\subsection{Existing spatial audio benchmarks}
\label{app:benchmarks}
Table~\ref{tab:benchmarks} compares MASU with existing spatial audio question-answering benchmarks. To the best of our knowledge, only the SO-Bench \cite{zhuSpatialOmniSpatialAudio2026} and ASA-Bench \cite{youWorldNotMono2026} benchmarks combine FOA input with more than a few question categories, which makes them the two candidates for an external evaluation. SO-Bench was part of the Spatial-Omni model work and scores its questions against a set of 63 class labels that are mostly derived from FSD50K \cite{fonsecaFSD50KOpenDataset2022}. These class labels result in coarse grouping of audio categories, while an audio captioning model is capable of more fine-grained description. However, the LLM-as-a-judge model would still rate the exact same event category as more correct than a more detailed description. The main paper therefore reports ASA-Bench, which also has its drawbacks. Since it is a 4-way multiple choice question benchmark, the default baseline would normally have a 25\% chance of being correct. However, option B is by far the most common answer, for 49.2\% of the items (Section~\ref{sec:exp}). Similarly, STARSS23-QA uses real FOA recordings but has five question types, and answering ``no'' to every yes/no question scores 83.4\% on that type. STAR-Bench, SpatialSoundQA and MMAU-Pro use stereo or binaural audio, and SAVVY-Bench needs video, is recorded with a microphone array on smart glasses and mixes egocentric and allocentric questions. None of these benchmarks asks follow-up questions.

\begin{table}[!htb]
\centering
\caption{Spatial audio question-answering benchmarks. Real gives the share of real recordings and Cat.\ the number of question categories, counting only the spatial subsets of STAR-Bench and MMAU-Pro.}
\label{tab:benchmarks}
\footnotesize
\setlength{\tabcolsep}{2.5pt}
\begin{tabular*}{\columnwidth}{@{\extracolsep{\fill}}lccrcr@{}}
\toprule
\textbf{Benchmark} & \textbf{FOA} & \textbf{Audio only} & \textbf{Real (\%)} & \textbf{Multi-turn} & \textbf{Cat.} \\
\midrule
STAR-Bench~\cite{liuStarBenchProbingDeep2026} & $\times$ & $\checkmark$ & 100 & $\times$ & 3 \\
SAVVY-Bench~\cite{chenSAVVYSpatialAwareness2026} & $\times$ & $\times$ & 100 & $\times$ & 4 \\
SpatialSoundQA~\cite{zhengBATLearningReason2024} & $\times$ & $\checkmark$ & 0 & $\times$ & 5 \\
MMAU-Pro~\cite{kumarMMAUProChallengingComprehensive2026} & $\times$ & $\checkmark$ & 100 & $\times$ & -- \\
SO-Bench~\cite{zhuSpatialOmniSpatialAudio2026} & $\checkmark$ & $\checkmark$ & mixed & $\times$ & 16 \\
ASA-Bench~\cite{youWorldNotMono2026} & $\checkmark$ & $\checkmark$ & 0 & $\times$ & 18 \\
STARSS23-QA~\cite{sudarsanamSpatialAudioUnderstanding2025} & $\checkmark$ & $\checkmark$ & 100 & $\times$ & 5 \\
\midrule
\textbf{MASU (ours)} & $\checkmark$ & $\checkmark$ & 17 & $\checkmark$ & 42 \\
\bottomrule
\end{tabular*}
\end{table}

\subsection{Describing the scene before answering}
\label{app:describe}
An earlier SciPhi-Chat model was trained on single-turn questions and on two-turn conversations that first describe the scene, both with a 50\% split. It was evaluated on 2,000 multiple-choice questions about simulated scenes in two ways. The first asks each question directly. The second first asks the model to describe everything it hears and then asks the same question, with the model's own description in context. Describing first raises accuracy from 43.1\% to 49.2\%, with the largest gains in counting and source identification. This result motivated the structured scene description that SciPhi-Chat generates once per conversation, before the first answer. 

\subsection{Use of the directional channels}
\label{app:channels}
To make sure our model really learns from all channels in FOA, a test checks that the omnidirectional $W$ channel (i.e., monaural) is not the only channel utilized. This was tested by answering 48 direction questions about single simulated sources, 16 each for left or right, front or back, and above or below. The questions are asked with the full recording or a silent recording. Controls keep the $W$ channel fixed and permute the directional channels $X$, $Y$ and $Z$, remove them, or take them from a source in the opposite direction. Accuracy drops from 79.2\% with full FOA to 39.6\% with permuted channels, 18.8\% with $W$ only, and 12.5\% with the opposite direction (Table~\ref{tab:channels}). With this we can conclude that the answers are based on directional channel information.

\begin{table}[!htb]
\centering
\caption{Direction accuracy (\%) of an earlier SciPhi-Chat model on 48 single-source questions, 16 per axis, with the full recording and four audio controls. Bold marks the highest value per column.}
\label{tab:channels}
\footnotesize
\setlength{\tabcolsep}{3pt}
\begin{tabular*}{\columnwidth}{@{\extracolsep{\fill}}lrrrr@{}}
\toprule
\textbf{Audio} & \textbf{Left/right} & \textbf{Front/back} & \textbf{Above/below} & \textbf{All} \\
\midrule
Full FOA & \textbf{93.8} & \textbf{87.5} & \textbf{56.3} & \textbf{79.2} \\
$X$, $Y$, $Z$ permuted & 50.0 & 43.8 & 25.0 & 39.6 \\
$W$ only & 50.0 & 0.0 & 6.3 & 18.8 \\
Opposite direction & 6.3 & 12.5 & 18.8 & 12.5 \\
Silence & 50.0 & 0.0 & 0.0 & 16.7 \\
\bottomrule
\end{tabular*}
\end{table}

\subsection{Tool use}
\label{app:tools}
\begin{figure}[!htb]
\centering
\begin{tikzpicture}[tool/.style={anchor=west,xshift=2pt,inner sep=0,font=\scriptsize},
  group/.style={xshift=-2.65cm,rotate=90,anchor=south,inner sep=0,align=center,font=\scriptsize\itshape}]
\begin{axis}[
  width=3.05cm, height=3.70cm, scale only axis,
  xmin=-25, xmax=75, ymin=-0.6, ymax=10.6, y dir=reverse,
  xtick={-20,0,20,40,60}, xmajorgrids, grid style={black!12},
  ytick={0,1,2,3,4,5,6,7,8,9,10},
  yticklabels={{Event class},{Count},{Elevation},{Azimuth},{Relative direction},{Location from identity},{Identity from location},{Scene description},{Localization},{Count},{Source binding}},
  ytick style={draw=none}, tick label style={font=\scriptsize},
  yticklabel style={text width=2.35cm,align=right},
  xlabel={$\leftarrow$ model better \quad Shapley value (pp) \quad tool better $\rightarrow$},
  xlabel style={font=\scriptsize}, clip=false,
  error bars/error bar style={line width=0.5pt},
  error bars/error mark options={rotate=90,mark size=1.4pt,line width=0.5pt}]
\draw[soft,line width=0.6pt] (axis cs:0,-0.6) -- (axis cs:0,10.6);
\node[group] at (axis cs:-25,3.0) {ASA-Bench};
\draw[black!25] (axis cs:-25,6.5) -- (axis cs:75,6.5);
\node[group] at (axis cs:-25,8.5) {STARSS23\\recordings};
\addplot+[only marks,accent,mark=*,mark options={fill=accent},mark size=1.5pt,line width=0.6pt,error bars/.cd,x dir=both,x explicit] coordinates {
  (44.9,0) -= (28.2,0) += (26.9,0)
  (30.1,1) -= (13.4,0) += (13.8,0)
  (0.9,4) -= (0.8,0) += (1.1,0)
};
\addplot+[only marks,accent,mark=*,mark options={fill=white},mark size=1.5pt,line width=0.6pt,error bars/.cd,x dir=both,x explicit] coordinates {
  (6.1,2) -= (6.4,0) += (6.9,0)
  (3.1,3) -= (3.3,0) += (4.5,0)
  (2.1,7) -= (2.1,0) += (3.0,0)
};
\addplot+[only marks,warm,mark=*,mark options={fill=white},mark size=1.5pt,line width=0.6pt,error bars/.cd,x dir=both,x explicit] coordinates {
  (-2.5,5) -= (3.1,0) += (2.8,0)
  (-2.7,6) -= (4.3,0) += (3.0,0)
  (-0.6,8) -= (3.1,0) += (2.6,0)
  (-0.7,9) -= (6.7,0) += (7.3,0)
  (-5.2,10) -= (8.8,0) += (10.0,0)
};
\node[tool] at (axis cs:75,0) {CLAP tagger};
\node[tool] at (axis cs:75,1) {DoA-cluster count};
\node[tool] at (axis cs:75,2) {STFT intensity DoA};
\node[tool] at (axis cs:75,3) {STFT intensity DoA};
\node[tool] at (axis cs:75,4) {STFT intensity DoA};
\node[tool] at (axis cs:75,5) {DRR distance};
\node[tool] at (axis cs:75,6) {Multi-window DoA};
\node[tool] at (axis cs:75,7) {BEATs tagger};
\node[tool] at (axis cs:75,8) {STFT intensity DoA};
\node[tool] at (axis cs:75,9) {DoA-cluster count};
\node[tool] at (axis cs:75,10) {BEATs tagger};
\end{axis}
\end{tikzpicture}
\caption{Shapley value (percentage points) of the best tool per task category for an earlier SciPhi-Chat model. Positive values favor the tool and negative values the model alone. Lines show 95\% bootstrap intervals and filled dots intervals that exclude zero. The best tool is chosen among tools that apply to at least half of the items in a category.}
\label{fig:shapley}
\end{figure}
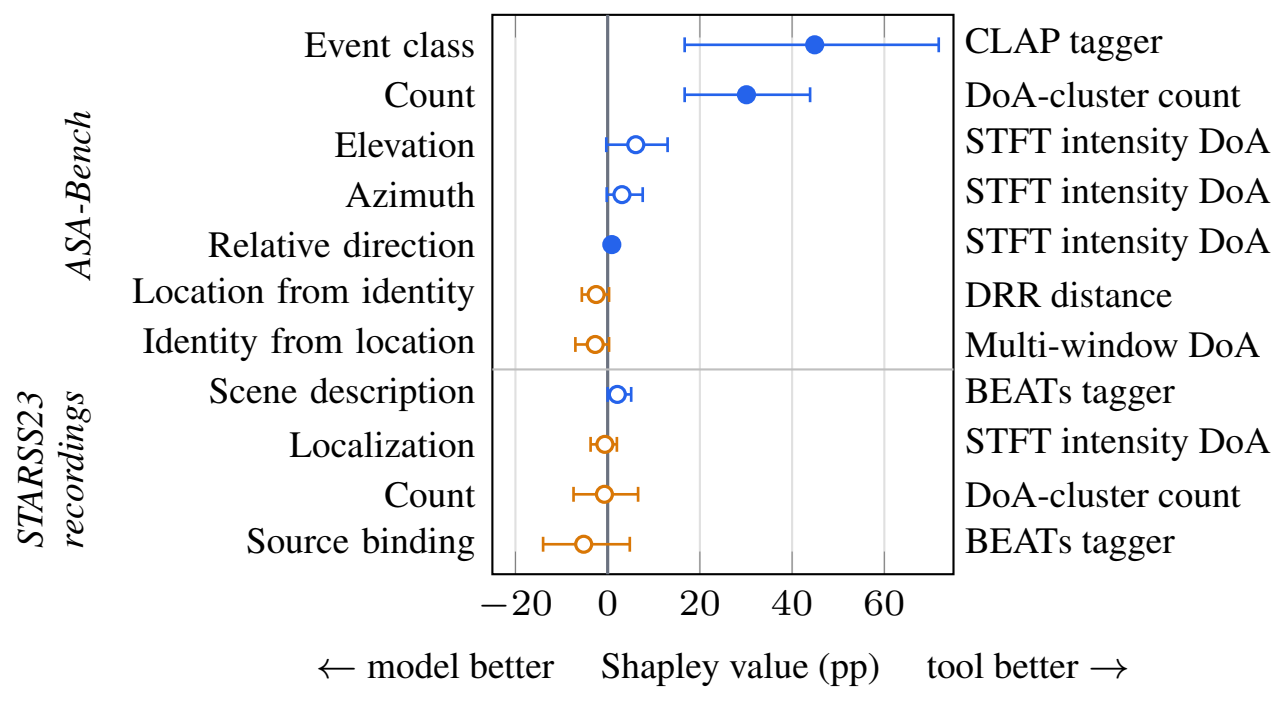

\noindent\textbf{Candidate tools.}
Besides the counting (DoA-cluster count), first localization (multi-window DoA), second localization (DPD DoA) and activity detection (onset span) tools of Section~\ref{sec:method}, 15 more tools were tested (see Figure~\ref{fig:shapley}). Intensity DoA and STFT intensity DoA compute the intensity direction over the whole recording or in time-frequency bins between 100 and 6000\,Hz, eigen-gap count counts sources from the covariance of the four channels, and DoA motion track follows the direction over time to tell moving from stationary sources. Steered beam measures the level of a virtual microphone pointed in one direction, DRR distance turns the energy ratio of the directional and $W$ channels into near, middle or far, and constant distance always answers 1.7\,m without using the audio. TF-mask separation and learned separation split the recording into one stream per source, by direction-based masking or with a 2.2M-parameter direction and CLAP-conditioned U-Net. The BEATs tagger \cite{chenBEATsAudioPreTraining2022}, the CLAP tagger \cite{elizaldeCLAPLearningAudio2023}, Whisper ASR \cite{radfordRobustSpeechRecognition2023}, and the DCASE 2024 SELD baseline model \cite{aparicioBaselineModelsEvaluation2024} name, transcribe, or detect and localize the sounds. Finally, the grounded solver computes relations such as ``more to the left'' from the outputs of the other tools, and self-invoke returns the model's own scene description as a second opinion.

\noindent\textbf{Which tools help which questions.}
An earlier SciPhi-Chat model was evaluated with all 19 tools. Following AudioToolAgent~\cite{wijngaardAudioToolAgentAgenticFramework2026}, the contribution of each tool per task category is its Shapley value over all tool combinations, measured against the model alone on 206 questions from ASA-Bench and STARSS23 recordings. Figure~\ref{fig:shapley} shows the best tool per category. Tools clearly help event classification and counting. For most other categories the interval includes zero, so the model alone answers them about as well, which motivates routing only selected categories.

\noindent\textbf{Routing between signal processing and the model.}
Signal processing answers different questions better than the model would. So a router can send each task to the more accurate of the two. An earlier SciPhi-Chat model was evaluated on the first 500 ASA-Bench items with such a router. It answers four tasks directly from intensity-vector estimates: the source count, the dominant source direction, that direction rotated by $180^\circ$, and, combined with a pretrained sound-event tagger, the source nearest to a queried direction. All other tasks keep the model answer. Routing raises accuracy from 59.4\% to 65.6\%. When choosing the routed tasks on one half of the ASA-Bench items and scoring the other half with the model, accuracy is 64.9\%. So the gain carries over to unseen items. On the first 500 SO-Bench items, routing the localization, distance and source-identification tasks, with pretrained audio-text models for source classes, raises accuracy from 27.5\% to 38.7\%.

\noindent\textbf{Tool calls during evaluation.}
The SciPhi-Chat model is capable of deciding on its own to call a specific tool, but performs well already without calling tools. Additionally, a further tool-trained variant on the earlier checkpoint trains on additional tool-calling conversations. Table~\ref{tab:tooluse} shows that it calls tools on more questions as training progresses, up to 15.7\%, while its MASU score neither improves nor decreases significantly. Of the calls that are not executed, 82.6\% name a tool that does not exist, 15.8\% have malformed arguments, and 1.6\% fail inside the tool. The share of executed calls rises from 46\% to about 70\% during training. 

Several issues prevented tool calling from improving accuracy. For example, in some cases where the model kept calling tools that fail, it ended up giving no answer to the question. Also, even though the model learned to use the tools, it often does not follow the tool output. For example, on 58 identification questions in which the tool output contained the correct answer, the earlier checkpoint still answered 70.7\% of them wrong. A possible cause is that in the supervised training data, the scene descriptions and tool outputs nearly always lead to the target answer, so the model learns to follow them. This can prevent the model from learning how to recover from its own errors. Techniques like reinforcement learning can be used in future work to improve the model's ability to call tools effectively. 

\begin{table}[!htb]
\centering
\caption{Tool calls of the tool-trained variant on the single-turn items of an earlier version of MASU. Called is the share of questions with a tool call and executed the share of calls that run without error. Steps count from the start of tool training. Bold marks the highest value per column.}
\label{tab:tooluse}
\footnotesize
\setlength{\tabcolsep}{3pt}
\begin{tabular*}{\columnwidth}{@{\extracolsep{\fill}}rrrr@{}}
\toprule
\textbf{Step} & \textbf{Called (\%)} & \textbf{Executed (\%)} & \textbf{MASU (\%)} \\
\midrule
376 & 3.8 & 46 & \textbf{45.1} \\
1,504 & 6.8 & 67 & 43.4 \\
2,632 & 11.8 & 60 & 43.3 \\
3,384 & 14.3 & \textbf{71} & 43.0 \\
3,760 & \textbf{15.7} & 68 & 44.1 \\
\bottomrule
\end{tabular*}
\end{table}

\end{document}